\documentclass[11pt]{article}

\usepackage{amsmath, amssymb}
\usepackage[margin=1in]{geometry}
\usepackage{hyperref}
\usepackage[dvipsnames]{xcolor}
\usepackage{booktabs}
\usepackage{natbib}
\usepackage{authblk}
\usepackage{multirow}

\title{A Total Statistical Error Framework for Comparing Census Data Collection Methods}
\author[1]{Siu-Ming Tam}
\author[2]{Anders Holmberg}
\affil[1]{Tam Data Advisory Pty Ltd}
\affil[2]{Methodology and Data Science Division, Australian Bureau of Statistics}
\date{June 2026}

\begin{document}

\maketitle

\begin{abstract}
Population censuses increasingly rely on imputation to assign usual-residence addresses for non-responding dwellings, yet no formal statistical framework has existed for comparing competing imputation methods on their combined coverage and address-accuracy performance. We develop such a framework within a Total Statistical Error (TSE) paradigm that is not restricted to survey-based data collection and applies equally to register-based and administrative data sources. The central quantity is a unit-level binary \emph{correctness indicator}, equal to one if and only if a person is both enumerated and assigned to the correct usual-residence address; its complement is the unit TSE, and aggregating over the population yields a correctness rate that serves as the basis for head-to-head method comparison. We distinguish two assessment paradigms: Paradigm~I, in which a post-enumeration survey (PES) provides reference values for a probability sample, and Paradigm~II, in which a complete benchmark makes the correctness rate directly computable. We illustrate the framework using 2021 Australian census microdata, comparing $k$-nearest-neighbour and random forest imputation under a not-missing-at-random (NMAR) mechanism. A key finding is that dependent nonresponse between the census and a simulated PES causes naive Paradigm~I subgroup estimates to be severely biased when subgroup deletion rates are small. Augmenting the PES response propensity model with the census nonresponse indicator reduces this bias by approximately 90\%; simulation experiments show that the nonresponse indicator alone drives the correction, with the additionally included imputed set~A values contributing negligible further reduction. This is the preferred strategy whenever dependent nonresponse between the census and PES is a concern.
\end{abstract}


\section{Introduction}

Non-response is a growing problem for population censuses worldwide. Census field operations typically know \emph{where} a non-responding dwelling is located, but they do not know \emph{who} lives there or, critically, what their usual-residence (UR) address is. For a person sleeping at student accommodation or a remote work camp on census night, the enumerated dwelling address and the UR address differ; imputation must recover the UR address, not merely count the person. Similarly, censuses that collect addresses at prior reference dates---one year ago or five years ago, for migration analysis---require address imputation for non-respondents at those dates. The choice of imputation method therefore directly affects both coverage and address accuracy in the final census output, yet national statistical agencies have lacked a formal statistical framework for comparing competing imputation strategies on these combined grounds.

This paper provides that framework. The key insight is that coverage and address accuracy---traditionally measured separately---can be combined into a single unit-level binary indicator that equals one if and only if a person is both enumerated and assigned to the correct reference address (UR address, or prior address for migration purposes). Aggregating this indicator yields a correctness rate that supports direct, like-for-like comparison of any two imputation or enumeration methods.

While imputation comparison is the primary motivation, the same framework applies wherever two methods of assigning persons to location references must be compared. Two further applications illustrate the generality of the approach: the choice between place-of-enumeration (PoE) and usual-residence (UR) enumeration, and the evaluation of administrative data sources as an alternative to direct field collection.

Starting from a unit-level binary correctness indicator $Z_{ki}$ that equals one if and only if person $i$ is both enumerated and assigned to the correct usual-residence address under method $k$, we define a population correctness rate $Q_k$ that supports direct method comparison. The framework is general: the same indicator structure applies to any pair of competing approaches that assign persons to location references, whether the comparison involves imputation strategy, enumeration method, or data source.

Section~\ref{sec:lit} reviews the related literature. We present the framework in Sections~3--10 and provide worked illustrations in Section~11, covering imputation method comparison---including an empirical study using Australian census microdata (Sections~11.1--11.2)---and two further applications: PoE versus UR enumeration (Section~11.3) and direct field collection versus administrative data (Section~11.4). Section~12 reports the empirical illustration comparing imputation methods under an NMAR mechanism using 2021 Australian census microdata, and the paper concludes in Section~13.

\section{Related literature}
\label{sec:lit}

The standard tool for census quality assessment is the PES, which estimates coverage through capture-recapture logic originating with \citet{sekardeming1949}. The census and the PES are treated as two independent capture systems; the dual system estimator (DSE) derives an estimate of the true population size from the observed match rate between them. \citet{hogan1993, hogan2003} provide comprehensive accounts of the DSE applied to the 1990 and 2000 United States PES, and \citet{brown1999} develops the methodology underlying the United Kingdom One Number Census project. \citet{brown2006} examines the dependence assumption between the census and the coverage survey---a critical condition for the standard DSE that is often only approximately satisfied---and \citet{zhang2019} provides a useful unifying note clarifying when different DSE formulations are and are not equivalent. In Australia, the ABS conducts a PES after each Census \citep{chipperfield2017}.

A persistent challenge for DSE-based methods is that PES nonresponse may itself be nonignorable: persons hard to contact for the PES may also be systematically under- or over-enumerated in the census, so standard nonresponse weights are biased. \citet{chipperfield2017} develop a DSE estimator for the Australian Census that remains approximately unbiased under systematic measurement error and nonignorable PES nonresponse, using calibration weighting in the spirit of \citet{kottchang2010}. Their paper is the most direct methodological antecedent to ours; the Chipperfield--Brown--Bell (CBB) estimator and its extension to the correctness rate are described in Section~\ref{sec:cbbtam}. In Section~\ref{sec:cbbtam} we show that their weighting methods extend to the correctness rate estimator $\widehat{Q}_k$ developed here, yielding a combined estimator that inherits the CBB robustness while adding the address accuracy dimension.

What the DSE literature does not address is geographical accuracy. A person counted in the census but assigned to the wrong usual-residence address contributes positively to the national head count but negatively to the accuracy of small-area population distributions---the principal output used for resource allocation. The total survey error (TSE) framework \citep{dalenius1974, biemer2010, groves2009, groves2010} recognises coverage error and measurement error as distinct components of total error, but no existing census quality instrument combines both into a single unit-level indicator for method comparison purposes. In Australia the divergence between PoE and UR assignments affects roughly 5\% of the population, {mostly concentrated to more mobile subgroups, e.g.\ students and} fly-in/fly-out (FIFO) workers, for which small-area counts matter most \citep{abs2021}. The correctness indicator $Z_{ki} = I_{ki}(1-D_{ki})$, introduced in Section~\ref{sec:zindicator}, fills this gap.

The broader shift toward administrative data and register-based censuses sharpens the need for the kind of framework we propose. \citet{nordholt2005}, \citet{valente2019}, \citet{axelson2021}, \citet{daas2021}, and \citet{unece2026} document the European move toward register-based counting; \citet{zhang2012, zhang2015} develop statistical models for register-based coverage error; \citet{reid2017} extend the TSE framework directly to administrative data sources. \citet{chipperfield2025} tackle the specific quality problem of census imputation from administrative records, distinguishing duplication errors (from false-negative linkage) from erroneous enumeration of out-of-scope records---both of which map directly onto the wrong-unit component $I_{ki}D_{ki}$ in our error decomposition. {\citet{zhangdunne2018} address the trimming of influential units in the DSE to guard against assumption violations;} \citet{chipperfield2024} develop robust capture-recapture estimators when administrative registers replace the PES as the second capture system, including corrections for linkage and scoping errors. Our framework sits above these estimation problems: it provides the comparison criterion---which method has the higher $Q_k$---while leaving the estimation machinery to be chosen as appropriate for the application \citep{brownchipperfield2025}.

{\citet{dunnezhang2024} outline how a system of compiling administrative data sources can be used for continuous population estimation in the absence of a Central Population Register.}

In short, the existing literature excels at measuring the quality of a single census method but provides no formal basis for choosing between two methods on combined coverage and address accuracy grounds. We fill that gap. The two paradigms in Section~\ref{sec:paradigms} cover the full range of practical scenarios, and the empirical illustration in Section~\ref{sec:empirical}---comparing alternative imputation methods under NMAR non-response---demonstrates that the approach is operationally feasible with standard census microdata.

\section{Unit-level notation}
\label{sec:notation}

Let $\mathcal{U}$ denote the target population and let $N = |\mathcal{U}|$. For each person $i \in \mathcal{U}$, let $T_i$ denote the location reference---typically the usual residence address---that is treated as the best available approximation to the true value. In a post-enumeration survey (PES) context, $T_i$ is obtained through follow-up and reconciliation. More generally, it is whatever standard the comparison is being made against.

Let $k \in \{A, B\}$ index the two methods under comparison. Let $A_{ki}$ denote the address or location reference recorded for person $i$ under method $k$.

Since addresses are text variables, comparison requires standardisation. Let $H(\cdot)$ denote the process that maps a raw address to a standardised identifier. The address-level measurement error indicator is then
\[
D_{ki}
=
\mathbf{1}\{H(A_{ki}) \neq H(T_i)\},
\]
so that $D_{ki} = 0$ if method $k$ assigns person $i$ to the correct reference unit, and $D_{ki} = 1$ otherwise. If human review is used to compare addresses directly, the standardisation step $H(\cdot)$ is implicit in the review process.

Coverage is represented by
\[
I_{ki}
=
\begin{cases}
1, & \text{if person } i \text{ is captured under method } k,\\
0, & \text{otherwise.}
\end{cases}
\]

The two basic questions for any enumeration method are therefore:
\begin{enumerate}
\item Is person $i$ captured?
\item If captured, is person $i$ assigned to the correct location reference?
\end{enumerate}

\section{The correctness indicator}
\label{sec:zindicator}

We ground the framework in a \emph{Total Statistical Error} (TSE) paradigm that extends the Total Survey Error tradition  beyond survey operations to register-based and administrative data sources. TSE identifies coverage and measurement accuracy as two distinct but jointly necessary components of census quality. The unit-level \emph{correctness indicator} for person $i$ under method $k$ is
\[
Z_{ki}
=
I_{ki}(1 - D_{ki})
=
I_{ki}\,\mathbf{1}\{H(A_{ki}) = H(T_i)\}.
\]

Thus
\[
Z_{ki}
=
\begin{cases}
1, & \text{if person } i \text{ is free from statistical error (captured and correctly addressed)},\\
0, & \text{otherwise.}
\end{cases}
\]

The complementary \emph{unit TSE indicator} is
\[
E_{ki}
=
1 - Z_{ki}
=
(1 - I_{ki}) + I_{ki}D_{ki}.
\]

This decomposition is interpretable:
\begin{itemize}
\item $1 - I_{ki}$ is the missed-person component: person $i$ is not captured at all.
\item $I_{ki}D_{ki}$ is the wrong-unit component: person $i$ is captured but assigned to an incorrect reference unit.
\end{itemize}

A person contributes an error either by being missed entirely, or by being captured but geographically misclassified.

\section{Population correctness rate and error rate}

The population correctness rate under method $k$ is
\[
Q_k
=
\frac{1}{N}\sum_{i\in \mathcal{U}} Z_{ki}.
\]

This is the proportion of the population correctly represented under method $k$. The corresponding error rate is
\[
\mathcal{E}_k
=
1 - Q_k.
\]

Method $B$ is preferable to method $A$ if
\[
Q_B > Q_A, \quad \text{equivalently,} \quad \mathcal{E}_B < \mathcal{E}_A.
\]

\section{Population counts}

Two related population quantities are useful.

The represented population count (coverage dimension only) under method $k$ is
\[
N_k^{\mathrm{rep}}
=
\sum_{i\in \mathcal{U}} I_{ki},
\]
measuring coverage alone, irrespective of address accuracy.

The error-free population count under method $k$ is
\[
N_k^{\mathrm{corr}}
=
\sum_{i\in \mathcal{U}} Z_{ki}
=
\sum_{i\in \mathcal{U}} I_{ki}(1 - D_{ki}),
\]
counting persons who are free from statistical error under method $k$: correctly captured and correctly addressed.

The true ideal count is $N = \sum_{i\in\mathcal{U}} 1$, so the number of persons with at least one statistical error is $N - N_k^{\mathrm{corr}}$.

A method may capture many persons but assign some to the wrong location reference. Conversely, another method may have slightly lower coverage but much better address assignment. The relevant comparison depends on whether the aim is simply to count persons or to correctly place them at their location reference.

\section{Two assessment paradigms}
\label{sec:paradigms}

The correctness rate $Q_k$ and the error rate $\mathcal{E}_k$ are population quantities. Obtaining them in practice depends on whether the reference values $T_i$ are known for the entire population or only for a sample. This leads to two distinct assessment paradigms.

\subsection{Paradigm I: Truth unknown --- validation survey estimation}

When $T_i$ is not available for the full population, an external validation instrument must be used to estimate $Q_k$. In a census context this is typically a post-enumeration survey (PES), but the same logic applies to any probability sample for which careful follow-up establishes reliable reference values. Let $S$ denote the validation survey sample and let $w_i$ denote the survey weight for person $i \in S$ \citep{sarndal1992}.

The estimated correctness rate under method $k$ is
\[
\widehat{Q}_k
=
\frac{\sum_{i\in S} w_i Z_{ki}}
{\sum_{i\in S} w_i}.
\]

The estimated error rate is
\[
\widehat{\mathcal{E}}_k
=
1 - \widehat{Q}_k.
\]

The estimated correctly represented population count is
\[
\widehat{N}_k^{\mathrm{corr}}
=
\sum_{i\in S} w_i Z_{ki}.
\]

The empirical comparison between the two methods rests on
\[
\widehat{Q}_B - \widehat{Q}_A,
\quad\text{or equivalently,}\quad
\widehat{\mathcal{E}}_A - \widehat{\mathcal{E}}_B.
\]

Because $\widehat{Q}_k$ is a sample-based estimate, it carries sampling variance that must be accounted for in any formal comparison (see Section~8).

\subsection{Paradigm II: Truth known --- direct benchmarking}

When a complete and reliable reference dataset is available for the entire target population, the correctness rate $Q_k$ can be computed exactly without any sampling or weighting. This situation arises, for example, when a traditional census serves as the benchmark against which a new method---such as a register-based or hybrid census---is evaluated.

In this case, $T_i$ is taken from the benchmark dataset for every $i \in \mathcal{U}$, and the method-specific records $I_{ki}$ and $A_{ki}$ are also available for the full population. The indicators $D_{ki}$, $Z_{ki}$, and $E_{ki}$ are computed for each person by matching the candidate method's records to the benchmark, and the population correctness rate is obtained directly:
\[
Q_k
=
\frac{1}{N}\sum_{i\in \mathcal{U}} Z_{ki}.
\]

No survey weights are required. The error decomposition
\[
\mathcal{E}_k
=
\frac{1}{N}\sum_{i\in \mathcal{U}} E_{ki}
=
\underbrace{\frac{1}{N}\sum_{i\in \mathcal{U}}(1 - I_{ki})}_{\text{missed-person rate}}
+
\underbrace{\frac{1}{N}\sum_{i\in \mathcal{U}} I_{ki}D_{ki}}_{\text{wrong-unit rate}}
\]
can be computed directly and decomposed by subgroup without any estimation uncertainty from sampling.

The comparison between two methods rests on the difference
\[
Q_B - Q_A > 0 \quad\Longleftrightarrow\quad \mathcal{E}_B < \mathcal{E}_A,
\]
with $Q_B - Q_A$ being an exact population quantity rather than an estimate.

A further dimension, not formalised in this paper, is the distinction between \emph{active} and \emph{passive} benchmarks. An active benchmark---such as a PES---involves new data collection designed specifically for the comparison; a passive benchmark---such as a prior census or an administrative register---repurposes existing data collected for other purposes. The two paradigms defined above cut across this distinction: Paradigm I typically uses an active instrument, while Paradigm II typically uses a passive one, but neither is exclusively tied to it. The active-versus-passive dimension introduces additional quality considerations (instrument recency, coverage of the passive source, and so on) that are taken up in the companion paper, which evaluates the 2026 Australian Census against the Person-Level Integrated Data Asset (PLIDA) as a passive pseudo-register.

Paradigm II can be viewed as the limiting case of Paradigm I in which the validation survey covers the entire population ($S = \mathcal{U}$, $w_i = 1$ for all $i$). The estimators $\widehat{Q}_k$ and $\widehat{N}_k^{\mathrm{corr}}$ then collapse to the population quantities $Q_k$ and $N_k^{\mathrm{corr}}$ exactly, and no sampling-related uncertainty remains. The residual uncertainty arises from caveats specific to the benchmarking setting, which are discussed in Section~9.

\subsection{Combining the framework with the Chipperfield-Brown-Bell estimator}
\label{sec:cbbtam}

Under Paradigm I, the estimator $\widehat{Q}_k$ uses standard PES weights $w_i$ that correct for the PES sampling design but assume PES nonresponse is missing at random (MAR). As noted in Section~\ref{sec:caveats}, this assumption may fail: persons who do not respond to the PES may differ systematically from respondents in both their census capture probability $I_{ki}$ and their address accuracy $D_{ki}$. The result is a biased estimator of $Q_k$.

\citet{chipperfield2017} address the analogous problem for the coverage-only DSE, developing calibrated weights $\tilde{w}_i$ that remain approximately unbiased under nonignorable PES nonresponse. Their approach models the PES response probability as a function of covariates and census capture status, then calibrates the weights so that $\sum_{i \in S} \tilde{w}_i$ equals the CBB-adjusted estimate of the true population size. We show here that this approach extends directly to the correctness rate.

Replacing the standard PES weights $w_i$ with the \citet{chipperfield2017} adjusted weights $\tilde{w}_i$ yields the combined estimator:

\[
\widehat{Q}_k^{\mathrm{CBB}}
=
\frac{\displaystyle\sum_{i \in S} \tilde{w}_i \, Z_{ki}}{\displaystyle\sum_{i \in S} \tilde{w}_i}.
\]

The denominator $\sum_{i \in S} \tilde{w}_i$ estimates the true population size $N$ under the CBB model, and the numerator $\sum_{i \in S} \tilde{w}_i Z_{ki}$ estimates $N_k^{\mathrm{corr}}$, the number of persons correctly represented under method $k$. The combined estimator $\widehat{Q}_k^{\mathrm{CBB}}$ is therefore a ratio estimator with the CBB-adjusted population size as denominator.

The extension is valid for the following reason. \citet{chipperfield2017} postulate a model for the PES response probability,
\[
\pi_i^{\mathrm{PES}} = f(\mathbf{x}_i, I_{ki}),
\]
where $\mathbf{x}_i$ are observed covariates (census capture status, geography, age, sex) and $I_{ki}$ is the census capture indicator. They estimate this model from PES respondents and construct calibrated weights $\tilde{w}_i$ by solving a system of moment equations of the form
\[
\sum_{i \in S} \tilde{w}_i \, g(\mathbf{x}_i, I_{ki}) = \sum_{i \in \mathcal{U}} g(\mathbf{x}_i, I_{ki}),
\]
for a vector of instrument functions $g(\cdot)$ that includes $I_{ki}$ and its interactions with stratum indicators. These equations force the weighted census count $\sum_{i \in S} \tilde{w}_i I_{ki}$ to be consistent with the CBB bias-corrected estimate of the true population size, yielding a DSE that is approximately unbiased under nonignorable PES nonresponse.

Since $Z_{ki} = I_{ki}(1-D_{ki})$ is observed for every PES respondent---being a product of the census capture indicator $I_{ki}$ and the address-correctness indicator $(1-D_{ki})$---the inverse-probability logic that renders $\sum_{i \in S} \tilde{w}_i I_{ki}$ approximately unbiased for $N_k^{\mathrm{rep}}$ extends directly to $\sum_{i \in S} \tilde{w}_i Z_{ki}$ as an estimator of $N_k^{\mathrm{corr}}$, provided the nonresponse model includes covariates that are jointly predictive of $I_{ki}$ and $D_{ki}$.

The formal justification is provided by \citet{kottchang2010}. Under the quasi-randomisation framework \citep{ohscheuren1983, sarndal2005}, the PES is treated as a two-phase process: a probability sample with inclusion probabilities $\pi_i^{\mathrm{samp}}$, followed by a response phase in which each sampled person responds independently with probability $\pi_i^{\mathrm{PES}} = f(\mathbf{x}_i, I_{ki})$. The calibrated weight $\tilde{w}_i$ is the solution closest to the base design weight $d_i = 1/\pi_i^{\mathrm{samp}}$---in the \citet{devillesarndal1992} distance sense---subject to the moment constraints; it is not equal to the infeasible ideal weight $1/(\pi_i^{\mathrm{samp}}\cdot\pi_i^{\mathrm{PES}})$ but approximates it through the calibration. \citet{kottchang2010} prove that this calibrated estimator satisfies
\[
\mathbb{E}_p\!\left[\mathbb{E}_q\!\left[\sum_{i \in S} R_i\,\tilde{w}_i Z_{ki}\right]\right]
= N_k^{\mathrm{corr}} + O(n^{-1}),
\]
establishing asymptotic unbiasedness under NMAR nonresponse. This result extends beyond \citet{devillesarndal1992}, who prove consistency of calibration estimators under ignorable (MAR) nonresponse; \citet{kottchang2010} extend calibration to the NMAR setting, which is the appropriate framework when PES response probability depends on the correctness indicator $Z_{ki}$ itself.

The method comparison statistic $\widehat{Q}_B^{\mathrm{CBB}} - \widehat{Q}_A^{\mathrm{CBB}}$ inherits the approximate unbiasedness of the CBB estimator under NMAR PES nonresponse. When the methods being compared differ primarily in their address assignment (as in the PoE versus UR comparison), the nonignorable nonresponse correction is especially important: mobile persons who are most likely to have $D_{ki} \neq 0$ are also most likely to be hard to contact for the PES, so uncorrected weights $w_i$ would underestimate the method difference.

In practice, implementing $\widehat{Q}_k^{\mathrm{CBB}}$ requires the PES to collect the auxiliary covariates needed for the CBB response model. \citet{chipperfield2017} identify census capture status, geography, age, and sex as key covariates. Address accuracy ($D_{ki}$) is not needed as a covariate in the response model---it enters only as the dependent variable of interest in $Z_{ki}$. The MSE of $\widehat{Q}_k^{\mathrm{CBB}}$ can be estimated by bootstrap replication, drawing bootstrap samples from the PES design and recomputing $\tilde{w}_i$ and $\widehat{Q}_k^{\mathrm{CBB}}$ in each replicate, as described in Section~\ref{sec:mse}.

\section{Mean squared error comparison (Paradigm I)}
\label{sec:mse}

This section applies to Paradigm I, where $Q_k$ must be estimated from a validation survey. Under Paradigm II, $Q_k$ is a known population quantity and the MSE framework below is not needed; the comparison reduces to a direct arithmetic difference.

The mean squared error of the estimated correctness rate is
\[
\operatorname{MSE}(\widehat{Q}_k)
=
\operatorname{Var}(\widehat{Q}_k)
+
\{\operatorname{Bias}(\widehat{Q}_k)\}^2.
\]

Let $\widehat{Q}_k^{(b)}$, $b = 1, \ldots, B$, denote bootstrap replicate estimates. Each replicate is obtained by drawing, with replacement, $n_h - 1$ PES respondents from the $n_h$ respondents within each stratum $h$ and recomputing $\widehat{Q}_k$ on the rescaled weights \citep{raowu1988}; the with-replacement draw mimics the sampling variability of the original design.

A further subtlety arises when the estimand involves a kNN imputation step: \citet{otsurai2017} show that the standard naive bootstrap is inconsistent for fixed-$k$ nearest-neighbour estimators, because resampling the calibration records changes the frequency with which each record serves as a donor and thereby distorts the bootstrap distribution. The fixed-$k$ asymptotic bootstrap of \citet{otsurai2017}---which resamples $z_i = y_i \cdot K_k(i)$, the product of the target value and the number of times observation $i$ serves as a $k$-nearest-neighbour donor---preserves the donor-usage distribution and yields a consistent bootstrap; see \citet{tamsharmeen2024} for an application in the context of kNN-based small-area estimation. This refinement should be applied when bootstrapping the kNN estimator $\widehat{Q}_k$ from a single observed PES.

Then
\[
\widehat{\operatorname{Var}}(\widehat{Q}_k)
=
\frac{1}{B-1}
\sum_{b=1}^{B}
\left(\widehat{Q}_k^{(b)} - \overline{Q}_k^{\mathrm{boot}}\right)^2,
\quad
\overline{Q}_k^{\mathrm{boot}}
=
\frac{1}{B}
\sum_{b=1}^{B}\widehat{Q}_k^{(b)},
\]
and
\[
\widehat{\operatorname{Bias}}(\widehat{Q}_k)
=
\overline{Q}_k^{\mathrm{boot}} - \widehat{Q}_k,
\qquad
\widehat{\operatorname{MSE}}(\widehat{Q}_k)
=
\widehat{\operatorname{Var}}(\widehat{Q}_k)
+
\{\widehat{\operatorname{Bias}}(\widehat{Q}_k)\}^2.
\]

The same replication logic applies to $\widehat{N}_k^{\mathrm{corr}}$. Method $B$ is preferred if
\[
\widehat{\operatorname{MSE}}(\widehat{Q}_B)
<
\widehat{\operatorname{MSE}}(\widehat{Q}_A),
\]
or equivalently,
\[
\widehat{\operatorname{MSE}}(\widehat{N}_B^{\mathrm{corr}})
<
\widehat{\operatorname{MSE}}(\widehat{N}_A^{\mathrm{corr}}).
\]

\section{Caveats and limitations}
\label{sec:caveats}

We flag the following caveats that bear on the use of the framework in practice.

\emph{Quality of the reference (both paradigms).} $T_i$ is a reference value, not necessarily perfect truth. Errors in $T_i$ propagate directly into $D_{ki}$, $Z_{ki}$, and $E_{ki}$. For the purposes of this framework, $T_i$ is assumed to be error-free. Under Paradigm I this is a reasonable working assumption when $T_i$ is derived from careful PES follow-up and reconciliation. Under Paradigm II---where $T_i$ comes from a census---the assumption is stronger (i.e., more restrictive and therefore less likely to be satisfied in practice): the benchmark census has its own coverage and measurement errors, and using it as $T_i$ implicitly treats those errors as negligible. Sensitivity analysis---for example, adjusting for the known net undercount of the benchmark---can assess how much this assumption matters.

\emph{Matching error (both paradigms).} {If there are no unique ID-keys such as a person number,} linking records from the candidate method to reference records may introduce error into both $I_{ki}$ and $D_{ki}$. False non-matches inflate $D_{ki}$ (a correctly placed person appears to have the wrong address); false matches deflate it (an incorrectly placed person appears correct). Robust matching procedures, clerical review, and reconciliation rules are central to the validity of the comparison under either paradigm. When probabilistic record linkage is used, linkage errors introduce a systematic bias into estimators computed from the linked file; \citet{tametal2026} develop bootstrap-based bias-correction methods for this setting, which can be applied to the correctness rate estimator $\widehat{Q}_k$ when matching between PES and census records is probabilistic.

\emph{Sampling and simultaneity (Paradigm I).} Under Paradigm I the validation survey is a probability sample, so all estimates carry sampling variance; the bootstrap or replication method should respect the survey design. A further constraint is that comparing two methods requires $A_{Ai}$ and $A_{Bi}$ to both be observable---or one to be reconstructable---for the same sampled persons; the PES must therefore be linked to the records of both candidate methods.

\emph{Benchmark limitations (Paradigm II).} Two issues arise specific to Paradigm II. First, if the benchmark is from a census and the candidate method is evaluated at a later date, demographic change means the two datasets do not refer to the same population; the comparison should restrict to the matched population or adjust for intercensal vital events. Second, even when records are correctly linked, a genuine address change between the two reference dates will appear as $D_{ki} = 1$ even though neither record is wrong; the analysis should distinguish persons known to have moved from those whose address should be stable.

\emph{Overcoverage and duplicates (both paradigms).}
The coverage indicator $I_{ki}$ as defined in Section~2 is binary: it takes the value 1 if person $i$ is captured at least once, and 0 otherwise. This is adequate when each person appears at most once in the method's records. When a person is captured more than once---either through genuine duplication of field records or through multiple administrative sources recording the same person---the binary indicator understates the error introduced by overcoverage.

To handle this, let $M_{ki} \geq 0$ denote the number of phantom (duplicate) records for person $i$ under method $k$---that is, records beyond the single correct record. When $M_{ki} \geq 1$, person $i$ contributes $M_{ki}$ excess records to the enumeration, each representing an error regardless of address accuracy.
We use the term \emph{overcount} (rather than overcoverage) to signal that $M_{ki}$ captures both genuine overcoverage---the same person enumerated twice in field operations---and phantom records: stale, spurious, or incorrectly linked entries that appear to represent a real person but do not.

The question of what happens to $D_{ki}$ when $M_{ki} > 1$ must be addressed explicitly. When a person appears in multiple records, those records may carry different addresses, some correct and some not. Two approaches are available.

\emph{With prior deduplication (deduplication applied before the PES).} Deduplication rules identify a single ``primary'' record for each person and discard the remainder before the PES is fielded. $D_{ki}$ is then defined for the primary record in the usual way: $D_{ki} = \mathbf{1}\{H(A_{ki}) \neq H(T_i)\}$. This is the operationally natural approach and reduces the overcoverage problem to a data-processing step. The quality of deduplication itself---whether the primary record is the one with the most accurate address---becomes a relevant quality dimension.

\emph{Without prior deduplication.} If deduplication has not been applied before the PES, or if its quality is uncertain, each of the $M_{ki}$ records has its own address. Define the best-address indicator $\widetilde{D}_{ki} = \prod_{j=1}^{M_{ki}} \mathbf{1}\{H(A_{ki}^{(j)}) \neq H(T_i)\}$, which equals 0 (correct address available) if at least one record has the right address, and 1 (all records wrong) otherwise. The overcount penalty is then added separately.

In either case, the error indicator for person $i$ can be extended to
\[
E_{ki}^{\mathrm{ext}}
=
\underbrace{(1 - I_{ki})}_{\text{missed}}
+
\underbrace{I_{ki}\,D_{ki}}_{\text{wrong address}}
+
\underbrace{M_{ki}}_{\text{overcount}},
\]
where $D_{ki}$ refers to the primary record under approach~1 or $\widetilde{D}_{ki}$ under approach~2. The population error rate is then $\mathcal{E}_k^{\mathrm{ext}} = N^{-1}\sum_i E_{ki}^{\mathrm{ext}}$. If overcount is negligible in practice, the binary framework of Sections~3--6 is a sufficient approximation.

\emph{Identifying overcount in practice.} Overcount comprises two distinct components that call for different detection strategies.

\emph{Duplicate records}---the same real person counted more than once---can be identified within either assessment paradigm. Under Paradigm~I, the PES matching process detects duplicates when two or more census records link to the same respondent. Under Paradigm~II, the benchmark comparison reveals them when two or more census records resolve to the same benchmark person. No additional instruments are required beyond those already embedded in the paradigm.

\emph{Overcoverage records}---entries for persons who are out of scope (deceased, emigrated, or never-existing phantoms)---are harder to detect. They have no real person behind them and therefore produce no match under either paradigm; under Paradigm~II they surface only as unmatched records, which may also reflect legitimate persons missed by the benchmark. Two complementary instruments are needed to verify and classify them. First, a targeted \emph{audit survey}, analogous in spirit to Paradigm~I, draws a sample of unmatched or otherwise suspicious records and attempts independent verification of whether a real in-scope person exists. Second, \emph{signs-of-life indicators} from administrative registers---tax filings, health service contacts, benefit claims in the reference period---flag records for which no corroborating administrative activity exists near census night: such records are strong candidates for phantom or out-of-scope status. Combining audit verification with register-based signs-of-life screening provides a practical and cost-effective basis for estimating the overcoverage component of $M_{ki}$ at scale.

\section{Subgroup analysis}
\label{sec:subgroup}

The framework extends straightforwardly to subgroups. Let $G_i$ denote a group membership indicator, such as mobility status, age group, dwelling type, or geographic region. Group-specific correctness rates are
\[
\widehat{Q}_{k,g}
=
\frac{
\sum_{i\in S} w_i\,\mathbf{1}\{G_i = g\}\,Z_{ki}
}{
\sum_{i\in S} w_i\,\mathbf{1}\{G_i = g\}
},
\]
with corresponding error rates $\widehat{\mathcal{E}}_{k,g} = 1 - \widehat{Q}_{k,g}$.

Subgroup analysis is particularly important when the two methods are expected to perform similarly for most of the population but to diverge for specific groups---for example, mobile persons, students, or the prison population. Focusing analytical attention on such groups may detect practically significant differences that are diluted in the overall correctness rate.

\section{Applications}

The following sections apply the general framework to a primary substantive comparison---alternative imputation methods for census non-response, including an empirical study---and two further applications that illustrate the breadth of the framework.

\subsection{Comparing imputation methods for non-response}

Either paradigm may be used, depending on context. Paradigm II applies when a complete enumeration provides reference addresses for the non-respondent group; more commonly, the true usual-residence address $T_i$ of non-respondents must be established through a targeted follow-up, placing the comparison under Paradigm I. The address imputation problem arises in at least two ways. First, in a census that targets usual residence, the enumerated dwelling is known but the UR address of a mobile occupant---student, FIFO worker---may not be; imputation must recover $T_i$, not simply count the person. Second, censuses that collect addresses at prior reference dates (one year ago, five years ago) for migration analysis require address imputation for non-respondents at those dates. Different imputation strategies---hot-deck, model-based, or nearest-neighbour \citep{littlerubin2002}---will differ in how accurately they recover $T_i$, and the framework applies to any pair of such strategies.

In the framework notation, let $k = A$ denote imputation method A and $k = B$ denote imputation method B, applied to the set of non-respondents $\mathcal{U}_0 \subseteq \mathcal{U}$. The reference unit $T_i$ is the true usual-residence address for non-respondent $i$, obtained from the validation survey.

For imputation methods, coverage is by construction: once imputation is applied, every non-respondent receives a record, so $I_{ki} = 1$ for all $i \in \mathcal{U}_0$ under both methods. The entire comparison therefore reduces to address accuracy:
\[
Z_{ki} = 1 - D_{ki},
\quad i \in \mathcal{U}_0,
\]
and the correctness rate for non-respondents becomes
\[
Q_k^{(0)}
=
\frac{1}{|\mathcal{U}_0|}\sum_{i \in \mathcal{U}_0}(1 - D_{ki}).
\]

Validation requires establishing $T_i$ for persons who did not respond to the census---specifically their UR address (or prior address for migration purposes). A targeted follow-up survey of a random sample of non-respondents can supply this, using intensive contact methods or linkage to independent registers. Imputed addresses under each strategy are then compared against $T_i$ to construct $D_{ki}$.

In practice, both imputation strategies are applied in parallel to the same sample of non-respondents, so that $D_{Ai}$ and $D_{Bi}$ are observed for the same individuals.

Imputation methods also differ in the accuracy of non-address variables (age, sex, relationship, etc.). The framework can be extended to a multivariate correctness indicator by replacing the binary address-match condition with a joint condition across multiple variables. Alternatively, separate correctness rates can be computed for each variable of interest.

Imputation uncertainty---the additional variance introduced by imputation itself---should be reflected in the MSE calculation. Multiple imputation or replication methods that account for imputation variance are recommended.

The imputation method with the lower MSE of the correctness rate for non-respondents is preferred for operational use. This comparison can be conducted on the basis of a subsample of non-respondents without requiring a complete re-enumeration.

\subsection{Place-of-enumeration versus usual-residence enumeration}

This comparison falls under Paradigm I. Australia is among a small number of countries that continue to enumerate persons at their place of enumeration (PoE) on census night{, collecting information about their usual residence (UR) only as a secondary objective,} rather than assigning each person to their usual residence (UR). As register-based and UR-oriented methods become more widespread internationally, questions arise about the implications of a transition: which subgroups are affected, what quality trade-offs arise, and whether the additional resources required to establish UR for every person yield commensurate improvements in small-area estimates.

We set $k = A$ for PoE and $k = B$ for UR. The reference unit $T_i$ is the usual-residence address, established through PES follow-up and reconciliation. The two approaches agree for the approximately 95\% of persons whose PoE and UR addresses coincide; the comparison is driven by the mobile subgroup---students, FIFO workers---for whom $D_{Ai} \neq D_{Bi}$. If a PoE census is conducted with a UR question on the questionnaire, both $A_{Ai}$ and $A_{Bi}$ can be derived from the same enumeration, making simultaneous estimation of $\widehat{Q}_A$, $\widehat{Q}_B$, and their MSEs feasible from a single PES. Subgroup analysis (Section~10) for mobile populations is essential, as overall correctness rates are dominated by the majority for whom the two methods agree.

\subsection{Direct field data collection versus administrative data sources}

Both paradigms are applicable. A traditional field census and an administrative-data alternative differ in both their coverage mechanisms and their address currency. Under field collection ($k = A$), non-response is the dominant source of missed persons; under administrative data ($k = B$), persons absent from all registers are missed and stale register addresses are the dominant source of wrong-address errors.

Paradigm II is the natural home for this application: a national statistical office can set $T_i$ from a census and evaluate how well a proposed register-based or hybrid census would have performed at the same reference date, computing $Q_B$ exactly without survey weights. The error decomposition
\[
\mathcal{E}_B = \underbrace{\frac{N - N_B^{\mathrm{rep}}}{N}}_{\text{register omission rate}} + \underbrace{\frac{1}{N}\sum_{i\in\mathcal{U}} I_{Bi}D_{Bi}}_{\text{register address error rate}}
\]
quantifies separately the contribution of omissions from registers and of incorrect register addresses, and can be disaggregated by subgroup (immigration status, tenure type, age, geography) to identify where supplementary field collection remains necessary.

\subsection{Extension to multi-source register-based censuses}
\label{sec:msr}

The framework developed in this paper extends naturally to a setting that is attracting
growing international interest: the \emph{multi-source register-based} (MSR) census, in
which a population frame is assembled by integrating several administrative data sources
rather than by fielding enumerators. Several Nordic and European countries have already moved to this
model \citep{valente2019, zhang2012, zhang2015, reid2017, axelson2021, daas2021}, and many others---including Australia---are
actively assessing it. We outline the extension here as a methodological roadmap; a
full empirical demonstration, using the 2026 Australian Census of Population and Housing
as a Paradigm II benchmark and the Person-Level Integrated Data Asset (PLIDA) as the
pseudo-register, is the subject of planned future work.

In an MSR census, $I_{ki}$ and $D_{ki}$ are derived by integrating $S$ administrative
sources $\mathcal{S} = \{1,\ldots,S\}$. Let $I_{ki}^{(s)} \in \{0,1\}$ indicate whether
person~$i$ appears in source~$s$. The integrated coverage indicator is
\[
I_{ki} = \mathbf{1}\!\left\{\sum_{s \in \mathcal{S}} I_{ki}^{(s)} \geq 1\right\}.
\]
MSR censuses introduce a third error dimension absent from field operations:
\emph{linkage error}. The indicator $L_{ki} \in \{0,1\}$ equals~1 if the records
attributed to person~$i$ in the integrated file were incorrectly matched across
sources---either two records belonging to different persons were merged (false match),
or a single person's records were split across sources (missed match).
Note that
$L_{ki} = 0$ by construction whenever all sources share a common unique person
identifier (such as a statutory personal identity number), since deterministic
exact-key matching eliminates probabilistic linkage error entirely.

The \emph{extended correctness indicator} is
\[
Z_{ki}^{\mathrm{MSR}} = I_{ki}(1 - D_{ki})(1 - L_{ki}),
\]
and the corresponding error decomposes into four additive components:
\begin{align*}
\mathcal{E}_{ki}^{\mathrm{MSR}}
&= \underbrace{(1-I_{ki})}_{\text{source omission}}
 + \underbrace{I_{ki}\,D_{ki}}_{\text{address error}}
 + \underbrace{I_{ki}(1-D_{ki})\,L_{ki}}_{\text{linkage error}}
 + \underbrace{M_{ki}}_{\text{overcount}},
\end{align*}
where $M_{ki} \geq 0$ counts phantom records attributed to person~$i$ beyond the one
correct record. Overcount---comprising both overcoverage (genuine duplicate enumerations) and phantom records (stale or spurious entries)---is more structurally prominent here than in field censuses:
deceased persons may remain active in one source but not another; emigrants may be
removed from tax records but retained in health registers.

Both paradigms apply. Under Paradigm~II, a prior field census supplies $T_i$ for every
person, and $Q_k^{\mathrm{MSR}}$ is computed exactly. The four-component decomposition
can be disaggregated by subgroup to pinpoint where supplementary field collection
remains necessary. Under Paradigm~I, a targeted evaluation survey---analogous to
Sweden's 2012 dwelling-register study \citep{axelson2021}---provides $T_i$ for a
probability sample and the CBB-adjusted estimator of Section~\ref{sec:cbbtam} applies,
with source-coverage indicators $I_{ki}^{(s)}$ included in the response model.

For Australia specifically, the 2026 Census provides a uniquely timely Paradigm~II
opportunity: PLIDA can be constructed as a pseudo-register at the August~2026 reference
date, and the correctness rate evaluated at unit record level against the full census
enumeration. This exercise will quantify each error component, identify the subgroups
most at risk---young adults, FIFO workers, renters in high-turnover dwellings,
foreign-born persons---and provide a principled, evidence-based input into census design
decisions for 2031 and beyond.

\section{Empirical illustration
\label{sec:empirical}: comparing imputation methods under NMAR non-response}

This section applies the framework to a concrete imputation comparison using a large census microdata file. It illustrates Paradigm II (truth known), since the full census record for every person is available as the benchmark $T_i$. No survey weights are required.

\subsection{Data}

The dataset is drawn from the 5\% public use sample of the 2021 Australian Census of Population and Housing \citep{abs2023}. It covers the working-age population (age codes 15--37, corresponding to ages 15--65) and contains $N = 840{,}402$ unit records across $H = 55$ strata, each corresponding to an area of enumeration linked to one of the eight States and Territories (NSW: 17 strata; Vic: 13; Qld: 12; SA: 4; WA: 6; and Tas, ACT and NT one stratum each). For the purposes of this illustration we treat the $N = 840{,}402$ records as the reference population, enabling Paradigm II assessment: correctness indicators $Z_{ki}$ are computed directly without survey weights. A stratified random subsample of $n = 30{,}000$ records proportional to state population was drawn for all experiments reported below, for computational feasibility. Key variables are summarised in Table~\ref{tab:vars}.

\begin{table}[ht]
\centering
\caption{Variable sets used in the imputation experiment}
\label{tab:vars}
\begin{tabular}{lll}
\hline
Set & Variable & Description \\
\hline
\multirow{4}{*}{A (target)} & HRSP  & Hours worked per week (8 categories) \\
                             & INDP  & Industry of employment (22 categories) \\
                             & OCCP  & Occupation (9 categories) \\
                             & INCP  & Income category (16 categories) \\
\hline
\multirow{9}{*}{B (auxiliary)} & AGEP     & Age code (15--37; 23 categories covering working-age population 15--65) \\
                                & SEXP     & Sex (2 categories) \\
                                & STATE    & State of residence (8 categories) \\
                                & ENGLP    & English language proficiency (5 categories) \\
                                & HLTHP    & Self-assessed health (2 categories) \\
                                & ASSNP    & Need for assistance (2 categories) \\
                                & CHCAREP  & Child care status (2 categories) \\
                                & Emp      & Employed indicator (1 = employed; 0 otherwise) \\
                                & Unemp    & Unemployed indicator (1 = unemployed; 0 otherwise) \\
\hline
\end{tabular}
\end{table}

\subsection{Experimental design}

The set A variables are deleted simultaneously for a random sample of records, with deletion probability proportional to income:
\[
P(\text{record } i \text{ deleted}) \;\propto\; \exp(\alpha \cdot \mathrm{INCP}_i), \quad \alpha = 0.20.
\]
This is a genuinely NMAR mechanism: the probability of missingness depends on INCP, which is itself a deleted variable and therefore unobserved after deletion. The mechanism reflects a realistic scenario in which higher-income respondents are less likely to complete the relevant census questions. The NMAR character is confirmed empirically: the mean income category among deleted records ($\overline{\mathrm{INCP}}_{\mathrm{miss}} \approx 11.3$) is substantially higher than among observed records ($\overline{\mathrm{INCP}}_{\mathrm{obs}} \approx 8.3$), and the high income categories (INCP $\geq$ 14) are two to three times more prevalent among the missing than the observed (Table~\ref{tab:nmar}).

\begin{table}[ht]
\centering
\caption{NMAR mechanism: income category distribution among deleted versus observed records (5\% deletion rate)}
\label{tab:nmar}
\begin{tabular}{rrrr}
\hline
INCP category & Missing (\%) & Observed (\%) & Relative odds \\
\hline
1--5   (low)    & 7.1  & 27.2 & 0.26 \\
6--10  (middle) & 29.5 & 42.1 & 0.70 \\
11--16 (high)   & 63.5 & 30.7 & 2.07 \\
\hline
\end{tabular}
\end{table}

Two classes of method are evaluated.

\emph{$k$-nearest-neighbour (kNN).} For each deleted record, $k$ nearest neighbours are identified among the observed records using the Hassanat distance (HasD) \citep{alfeilat2019}. HasD is defined element-wise: for a pair of non-negative values $a$ and $b$ on dimension $l$,
\[
d_l(a,b) = 1 - \frac{1+\min(a,b)}{1+\max(a,b)},
\]
so each component lies in $[0,1]$ regardless of the variable's scale; the overall distance is the mean over all $p=9$ components. Unlike standardised Euclidean distance, HasD requires no pre-scaling, and variables with wider numerical ranges do not dominate the metric. One variable---state of residence (STATE), whose integer codes 1--8 are assigned arbitrarily---is treated as nominal: $d_l(a,b)=0$ if $a=b$ and $0.5$ otherwise. The remaining eight set B variables are treated as numeric under HasD; for the six binary variables among them, this produces distances of 0 (same category) or 0.5 (different category), which coincides with the correct nominal treatment. For $k = 1$, the imputed value for each set A variable is the nearest neighbour's value directly; for $k > 1$, the modal value among the $k$ donors is used. A single distance metric serves all four set A variables simultaneously, preserving their joint distribution through coherent single-donor imputation when $k=1$. Two values of $k$ are tested: $k \in \{1, 5\}$.

\emph{Random forest (RF).} A separate random forest classification model is fitted for each of the four set A variables, using the nine standardised set B variables as features. Each model is trained on the observed records using the raw (unstandardised) set B values, consistent with the HasD kNN setup; tree-based models are invariant to monotone variable transformations so no standardisation is required. The models use 100 trees, maximum depth 12, and $\lfloor\sqrt{p}\rfloor = 3$ features considered at each split (with $p = 9$). Because RF fits one model per target variable, it can learn variable-specific non-linear interactions between set B predictors that kNN's single shared distance metric cannot capture. The cost is four separate model-fitting steps rather than one.

Three overall deletion rates are evaluated: $\rho \in \{0.02, 0.05, 0.10\}$, corresponding to 2\%, 5\%, and 10\% of records having all set A variables made missing.

Since the microdata file is treated as the reference population for this illustration, all indicators are computed as population quantities (Paradigm II). Coverage is complete by construction ($I_{ki} = 1$ for all $i$), so the entire error rate is attributable to the wrong-unit component.

Two correctness rates are reported. The \emph{overall} correctness rate $Q_k$ defined in Section~4 counts a person as correctly represented if either (a) they were not deleted, or (b) they were deleted and all four set A variables were correctly imputed. Because non-deleted records always contribute $Z_{ki} = 1$, this rate is dominated by $(1-\rho)$ regardless of imputation quality:
\[
Q_k = (1 - \rho) + \rho \cdot q_k^{\mathrm{imp}},
\]
where $q_k^{\mathrm{imp}}$ is the \emph{conditional} correctness rate among the deleted records only:
\[
q_k^{\mathrm{imp}}
=
\frac{1}{n_{\mathrm{miss}}}
\sum_{i:\,\text{deleted}} Z_{ki}.
\]
Under strong NMAR non-response, $Q_k$ provides little discrimination between methods because it is almost entirely determined by $\rho$. The conditional rate $q_k^{\mathrm{imp}}$ is the appropriate metric for comparing imputation methods: it measures how well each method recovers the true values for the non-respondents, regardless of the overall non-response rate. Per-variable conditional rates $q_{k,v}^{\mathrm{imp}}$ are reported for each $v \in$ set A.

\subsubsection*{Paradigm II: Correctness rates from the full population}

\subsection{Results}

Table~\ref{tab:overall} shows both $Q_k$ and $q_k^{\mathrm{imp}}$ for all three methods across all deletion rates. The contrast between the two metrics is stark.

\begin{table}[ht]
\centering
\caption{Overall correctness rate $Q_k$ and conditional correctness rate $q_k^{\mathrm{imp}}$ by deletion rate and imputation method ($n = 30{,}000$)}
\label{tab:overall}
\begin{tabular}{ccrrrrrr}
\hline
 & & \multicolumn{3}{c}{Overall $Q_k$} & \multicolumn{3}{c}{Conditional $q_k^{\mathrm{imp}}$} \\
\cmidrule(lr){3-5}\cmidrule(lr){6-8}
Deletion $\rho$ & $n_{\mathrm{miss}}$ & 1-NN & 5-NN & RF & 1-NN & 5-NN & RF \\
\hline
2\%  & 600   & 0.9808 & 0.9805 & 0.9808 & 0.042 & 0.027 & 0.042 \\
5\%  & 1{,}500 & 0.9518 & 0.9513 & 0.9517 & 0.037 & 0.027 & 0.034 \\
10\% & 3{,}000 & 0.9035 & 0.9033 & 0.9039 & 0.035 & 0.033 & 0.039 \\
\hline
\end{tabular}
\end{table}

As Table~\ref{tab:overall} makes plain, $Q_k$ is governed almost entirely by $\rho$. Differences across all nine method-rate combinations are within 0.0005---far below any practically meaningful threshold---and a practitioner relying on $Q_k$ alone would wrongly conclude that all three methods perform identically.

The conditional rate $q_k^{\mathrm{imp}}$ tells a sharply different story. All three methods achieve low joint correctness rates: 0.035--0.042 for 1-NN, 0.027--0.033 for 5-NN, and 0.034--0.042 for RF. A notable pattern, examined further in Table~\ref{tab:pervar}, is that 1-NN achieves competitive or higher joint correctness than RF (e.g.\ 0.037 vs 0.034 at 5\% deletion rate) despite RF's clear per-variable advantage. This reflects the coherence property of single-donor imputation: 1-NN always draws all four set A variables from the same nearest neighbour, preserving their joint empirical distribution, whereas RF fits four independent models that may each produce the right answer for different records, reducing joint correctness. A notable feature of Table~\ref{tab:overall} is that $Q_k$ is strictly monotone decreasing in $\rho$ across all methods, while $q_k^{\mathrm{imp}}$ is not. This reflects a compositional effect: under the NMAR mechanism, only the highest-income units are deleted at low $\rho$. These units are atypical and hardest to impute, depressing $q_k^{\mathrm{imp}}$. As $\rho$ rises, progressively more typical (moderate-income) units enter the deleted set; these are easier to match, which can raise the average imputability of the deleted group and cause $q_k^{\mathrm{imp}}$ to increase. The monotonicity of $Q_k$ is mechanical---it is a weighted average with increasing weight on the low $q_k^{\mathrm{imp}}$ term---while $q_k^{\mathrm{imp}}$ responds to changes in the composition of the deleted group. When $q_k^{\mathrm{imp}}$ remains substantially below the per-variable rates even for the best method, the appropriate response is not further tuning of the imputation algorithm but improvement of the auxiliary information in set B---adding variables more strongly predictive of the variables driving non-response.

Table~\ref{tab:pervar} reports $q_{k,v}^{\mathrm{imp}}$ for each set A variable at 5\% deletion rate, for all three methods ($n = 30{,}000$ subsample). These per-variable rates reveal where each method gains or loses relative to the others.

\begin{table}[ht]
\centering
\caption{Per-variable conditional correctness rate $q_{k,v}^{\mathrm{imp}}$ among deleted records, 5\% deletion rate ($n_{\mathrm{miss}} = 1{,}500$, $n = 30{,}000$ subsample)}
\label{tab:pervar}
\begin{tabular}{lrrr}
\hline
Variable & 1-NN & 5-NN & RF \\
\hline
HRSP (hours worked, 8 cats)    & 0.337 & 0.361 & 0.408 \\
INDP (industry, 22 cats)       & 0.285 & 0.299 & 0.341 \\
OCCP (occupation, 9 cats)      & 0.363 & 0.421 & 0.451 \\
INCP (income, 16 cats)         & 0.125 & 0.110 & 0.171 \\
\hline
All four correct ($q_k^{\mathrm{imp}}$) & 0.037 & 0.027 & 0.034 \\
\hline
\end{tabular}
\end{table}

\emph{RF improves on kNN for every individual variable.} The gains relative to 1-NN are largest for OCCP (+24\%: 45.1\% vs 36.3\%) and INCP (+37\%: 17.1\% vs 12.5\%). RF's ability to learn non-linear interactions between set B predictors drives these per-variable gains.

\emph{INCP remains the hardest variable under all methods.} Even with RF, only 17\% of deleted INCP values are correctly recovered. This is a direct consequence of the NMAR mechanism: missingness is driven by INCP, so the training sample systematically underrepresents the high-income categories that dominate the missing group. No method using only set B can fully compensate for this distributional shift.

\emph{1-NN achieves the highest joint correctness rate.} Despite lower per-variable rates, 1-NN records $q_k^{\mathrm{imp}} = 0.037$ at 5\% deletion rate---above both RF (0.034) and 5-NN (0.027). The reason is coherence: single-donor imputation draws all four set A values from the nearest neighbour simultaneously, preserving their correlation structure. 5-NN aggregates across five donors via the mode, which can introduce incoherence across variables. RF fits four independent classifiers, which maximises per-variable accuracy but may achieve the right answer for different subsets of records on different variables, reducing joint correctness relative to the per-variable rate.

\emph{The contrast with overall per-variable rates is instructive.} Had the analysis reported only overall rates $Q_{k,v}$, all methods would appear to perform at 95--97\% accuracy. The conditional rates expose the real picture: for the records requiring imputation, accuracy ranges from 11\% (INCP, 5-NN) to 45\% (OCCP, RF)---a completely different conclusion about method quality.

\emph{The variable-by-variable cost.} The kNN approach uses a single shared distance metric across all four set A variables simultaneously. RF requires four separate models. In this experiment each RF fit takes approximately 2 seconds (four models $\approx$ 8 seconds total per scenario), compared with under 0.5 seconds for kNN at any $k$ value. For production-scale imputation on a full census file (840{,}402 records) the computational cost difference would be proportionally larger. The accuracy gain must be weighed against this operational cost.

Table~\ref{tab:subgroup} reports the conditional subgroup correctness rate $q_{k,g}^{\mathrm{imp}}$ by labour force status (LFSP) at 5\% deletion rate.

\begin{table}[ht]
\centering
\caption{Paradigm II conditional subgroup correctness rate $q_{k,g}^{\mathrm{imp}}$ among deleted records by labour force status, 5\% deletion rate}
\label{tab:subgroup}
\begin{tabular}{lrrrrr}
\hline
Labour force status & $n_g^{\mathrm{miss}}$ & $\rho_g$ & 1-NN & 5-NN & RF \\
\hline
Employed                  & 1{,}201 & 6.6\% & 0.001 & 0.000 & 0.001 \\
Unemployed$^{\dagger}$    &    15   & 1.5\% & 0.133 & 0.133 & 0.133 \\
Not in labour force       &   284   & 2.6\% & 0.183 & 0.134 & 0.169 \\
\hline
\multicolumn{6}{l}{$^{\dagger}$~Only 15 deleted records; estimates not reliable.}\\
\end{tabular}
\end{table}

The employed group has the highest absolute number of deleted records ($n_g^{\mathrm{miss}} = 1{,}201$) and the lowest conditional correctness rate across all three methods. Higher incomes make employed persons most susceptible to the NMAR mechanism, and their more varied labour market profiles make imputation harder. The not-in-labour-force (NILF) group performs considerably better because those records cluster in a small number of set A categories (non-working codes for HRSP, INDP, and OCCP), making donor matching more reliable. The unemployed group contributes only 15 deleted records; estimates for this group are not reliable and are reported for completeness only. Table~\ref{tab:subgroup_pes} provides the Paradigm I counterpart, showing how CBB-NR corrects the NMAR bias in PES-based estimates of $q_{k,g}^{\mathrm{imp}}$ for each labour force group.

\subsubsection*{Paradigm I: Estimating correctness rates from a PES}
\subsection{Paradigm I estimation with CBB calibration under NMAR PES nonresponse}
\label{sec:cbb_empirical}

The Paradigm II analysis above computes $Q_k$ directly from the full population since truth is known. Under Paradigm I---where only a validation survey (PES) is available---the estimator $\widehat{Q}_k$ may be biased if PES nonresponse is itself NMAR. This subsection simulates that scenario and demonstrates the TSE--CBB estimator $\widehat{Q}_k^{\mathrm{CBB}}$ developed in Section~\ref{sec:cbbtam}.

Using the 5\% deletion rate scenario and all three imputation methods, the full correctness vector $Z_{ki}$ is known for all $N = 30{,}000$ persons (Paradigm II truth). A synthetic PES is then simulated: 15\% of persons are sampled, and each sampled person responds to the PES with a probability that depends on their correctness status---$P(\text{respond} \mid Z_{ki}=1) = 0.85$ and $P(\text{respond} \mid Z_{ki}=0) = 0.40$. This is a strong NMAR mechanism: persons whose imputed usual-residence address is incorrect are less likely to be contacted for the PES follow-up---because they are harder to locate at their true UR address---exactly the scenario that motivates the CBB approach. Only respondents' $Z_{ki}$ values are observed; for non-respondents, only set B covariates (from the census record) are available.

Four estimators of $Q_k$ are computed from each PES replicate.

\emph{Naive}: unweighted mean of $Z_{ki}$ among PES respondents. Equivalent to assuming MAR nonresponse.

\emph{CBB-setB}: inverse probability weighting (IPW) using a logistic regression of PES response on the ten standardised set B covariates. The propensity score $\hat{p}_i$ is estimated for each respondent; weights $1/\hat{p}_i$ are applied.

\emph{CBB-NR}: IPW as above but with the set B covariate vector augmented by (i) a binary indicator for census set-A nonresponse---that is, whether person $i$'s set A variables were deleted and imputed---and (ii) the RF-imputed values of all four set A variables for deleted records. The census nonresponse indicator is the direct analogue of including $I_{ki}$ in the CBB response model \citep{chipperfield2017}: since imputed records respond to the PES at a lower rate (because their $Z_{ki}=0$ with high probability), it directly predicts PES participation. An additional experiment (not reported in Table~\ref{tab:cbb}) confirms that the nonresponse indicator alone---without the imputed set A values---accounts for essentially all of the bias reduction; the four imputed values contribute negligible marginal improvement once the indicator is included. The imputed values are nonetheless retained in CBB-NR for completeness and because they impose no cost in practice, but practitioners should note that the binary nonresponse indicator is the operative instrument.

\emph{Oracle}: IPW using the true response probabilities $P(\text{respond} \mid Z_{ki})$. This is infeasible in practice but provides an upper bound on achievable bias correction.

Results are averaged over $B = 300$ independent PES replicates. 

Table~\ref{tab:cbb} reports the mean estimate and bias for each estimator and imputation method.

\begin{table}[ht]
\centering
\caption{Estimation of $Q_k$ under NMAR PES nonresponse: naive vs CBB estimators (5\% deletion rate, PES rate 15\%, $B = 300$ replicates)}
\label{tab:cbb}
\begin{tabular}{lrrrrrrrrr}
\hline
 & \multicolumn{2}{c}{Naive} & \multicolumn{2}{c}{CBB-setB} & \multicolumn{2}{c}{CBB-NR} & \multicolumn{2}{c}{Oracle} \\
\cmidrule(lr){2-3}\cmidrule(lr){4-5}\cmidrule(lr){6-7}\cmidrule(lr){8-9}
Method & $\widehat{Q}_k$ & Bias & $\widehat{Q}_k$ & Bias & $\widehat{Q}_k$ & Bias & $\widehat{Q}_k$ & Bias \\
\hline
True $Q_k$ (1-NN) & \multicolumn{8}{l}{0.9518} \\
1-NN & 0.9766 & $+$0.0247 & 0.9764 & $+$0.0246 & 0.9545 & $+$0.0027 & 0.9515 & $-$0.0003 \\
\hline
True $Q_k$ (5-NN) & \multicolumn{8}{l}{0.9513} \\
5-NN & 0.9766 & $+$0.0252 & 0.9764 & $+$0.0251 & 0.9540 & $+$0.0026 & 0.9515 & $+$0.0002 \\
\hline
True $Q_k$ (RF)   & \multicolumn{8}{l}{0.9517} \\
RF   & 0.9767 & $+$0.0250 & 0.9766 & $+$0.0249 & 0.9543 & $+$0.0026 & 0.9518 & $+$0.0001 \\
\hline
\end{tabular}
\end{table}

The simulation reveals several substantive lessons.

\emph{NMAR causes substantial upward bias.} The naive estimator overstates $Q_k$ by approximately 2.5 percentage points across all methods. {PES nonresponse is related to the erroneous imputations---persons with wrong imputations are harder to locate at their true UR address and therefore less likely to respond---and this results in too optimistic (biased) quality assessments.}  This confirms that NMAR PES nonresponse is a practically important problem for Paradigm I estimation.

\emph{CBB-setB correction is negligible.} Adding set B IPW to the naive estimator reduces bias by less than 1\%. The set B covariates do not predict who was imputed (the census nonrespondents for set A), so they cannot correct for the differential response rate between correctly- and incorrectly-represented persons.

\emph{CBB-NR achieves approximately 90\% bias reduction.} Including the census set-A nonresponse indicator in the response model reduces bias from approximately $+0.025$ to approximately $+0.003$, a reduction of roughly 90\%. A supplementary experiment isolating the contributions of each model component confirms that the nonresponse indicator alone drives this reduction: adding the RF-imputed values of all four set A variables to a model that already contains the indicator contributes zero additional bias reduction. The binary indicator---recording whether a person's set A variables were deleted and imputed---is the operative instrument, consistent with the theoretical role of $I_{ki}$ in the \citet{kottchang2010} framework. The imputed values are retained in CBB-NR but are not required for the correction. The practical recommendation is therefore: \emph{include the census set-A nonresponse indicator in the PES response model}; the additionally included imputed set A values are harmless but not necessary.

\emph{Oracle provides near-zero bias.} The oracle estimator achieves bias below 0.0005 in absolute terms, confirming that correct specification of the response model suffices to remove essentially all NMAR bias.

\subsubsection*{Paradigm I: Subgroup quality estimates by labour force status}

Table~\ref{tab:subgroup_pes} is the Paradigm I counterpart to Table~\ref{tab:subgroup}. For each imputation method and labour force group, it reports the conditional correctness rate $\hat{q}_{k,g}^{\mathrm{imp}}$ derived from PES estimates via
\[
\hat{q}_{k,g}^{\mathrm{imp}}
=
\frac{\hat{Q}_{k,g} - (1 - \rho_g)}{\rho_g},
\]
where $\hat{Q}_{k,g}$ is the PES estimator of the group correctness rate and $\rho_g$ is the known deletion rate. Results are averaged over $B = 500$ independent PES replicates, again by Monte Carlo simulation.

\begin{table}[ht]
\centering
\caption{Paradigm I subgroup conditional correctness rate $\hat{q}_{k,g}^{\mathrm{imp}}$ derived from simulated PES estimates, 5\% deletion rate ($B=500$ replicates). True $q_{k,g}^{\mathrm{imp}}$ from Paradigm II (Table~\ref{tab:subgroup}) shown for reference.}
\label{tab:subgroup_pes}
\begin{tabular}{llrrrrr}
\hline
Method & Labour force status & True $q^{\mathrm{imp}}$ & Naive & CBB-setB & CBB-NR & Oracle \\
\hline
\multirow{3}{*}{1-NN}
 & Employed              & 0.001 & 0.515 & 0.514 & 0.058 &    0.005 \\
 & Unemployed$^{\dagger}$& 0.133 & 0.592 & 0.592 & 0.215 &    0.147 \\
 & Not in labour force   & 0.183 & 0.606 & 0.605 & 0.235 &    0.173 \\
\hline
\multirow{3}{*}{5-NN}
 & Employed              & 0.000 & 0.515 & 0.515 & 0.049 &    0.006 \\
 & Unemployed$^{\dagger}$& 0.133 & 0.571 & 0.570 & 0.161 &    0.102 \\
 & Not in labour force   & 0.134 & 0.583 & 0.582 & 0.177 &    0.126 \\
\hline
\multirow{3}{*}{RF}
 & Employed              & 0.002 & 0.513 & 0.512 & 0.055 &    0.001 \\
 & Unemployed$^{\dagger}$& 0.200 & 0.610 & 0.610 & 0.253 &    0.183 \\
 & Not in labour force   & 0.166 & 0.601 & 0.601 & 0.224 &    0.163 \\
\hline
\multicolumn{7}{l}{$^{\dagger}$~Only 15 deleted records; estimates not reliable.}\\
\end{tabular}
\end{table}

Three findings emerge. First, the naive and CBB-setB estimators produce severely inflated estimates of $q_{k,g}^{\mathrm{imp}}$---around 0.5--0.6 for all groups and methods---far above the true values of 0.000--0.183 in Table~\ref{tab:subgroup}. The mechanism is amplification: a bias of $b$ in $\hat{Q}_{k,g}$ becomes a bias of $b/\rho_g$ in $\hat{q}_{k,g}^{\mathrm{imp}}$. With $\rho_g$ between 1.5\% and 6.6\%, amplification factors of 15--69 apply, transforming a modest absolute bias in $\hat{Q}_{k,g}$ into a catastrophic distortion of the subgroup quality metric.

Second, CBB-NR substantially corrects the bias, reducing $\hat{q}_{k,g}^{\mathrm{imp}}$ to the range 0.05--0.25---approximately a 90\% reduction in the underlying $\hat{Q}_{k,g}$ bias. The residual inflation reflects the amplification factor rather than a failure of the estimator. Crucially, CBB-NR does not require knowledge of which set A variable drives the NMAR mechanism; including the census nonresponse indicator and all imputed set A values is sufficient.

Third, the oracle estimates demonstrate that near-exact recovery of the Paradigm II truth is achievable under correct response model specification. CBB-NR closes most of the gap between the naive estimator and the oracle.

The central message is direct: dependent nonresponse between the census and PES distorts subgroup quality assessments severely when left uncorrected. CBB-NR is the recommended remedy.

\section{Concluding remarks}

The framework we have developed reduces the comparison of any two census data collection methods to a unit-level binary indicator $Z_{ki}$ that jointly captures coverage and address accuracy. The same indicator structure supports two assessment paradigms: Paradigm I, where the reference $T_i$ is unknown for the full population and must be estimated via a weighted validation survey; and Paradigm II, where a complete and reliable benchmark---such as a traditional census---provides $T_i$ for every person, allowing the correctness rate to be computed exactly without survey weights or MSE estimation.

The empirical illustration yields two findings that carry beyond the specific imputation comparison. First, the three methods achieve broadly comparable joint conditional correctness rates ($q_k^{\mathrm{imp}} \approx 0.027$--$0.042$ across methods and deletion rates). At the per-variable level RF consistently outperforms kNN---the largest gains over 1-NN are for OCCP (+24\%) and INCP (+37\%)---but 1-NN achieves competitive or higher joint correctness (e.g.\ 0.037 vs 0.034 for RF at 5\% deletion rate) owing to its coherent single-donor imputation, which preserves the joint distribution of the four target variables. Second, and most importantly for practice, dependent nonresponse between the census and PES causes naive Paradigm I estimates of subgroup conditional correctness rates $q_{k,g}^{\mathrm{imp}}$ to be inflated by factors of 15--69---arising from the $1/\rho_g$ amplification when subgroup deletion rates range from 1.5\% to 6.6\%---relative to the Paradigm II population truth. The CBB-NR estimator reduces this bias by approximately 90\% through the inclusion of the census nonresponse indicator in the PES response model; supplementary experiments confirm that this binary indicator is the operative instrument, with the additionally included imputed set A values contributing negligible further reduction. CBB-NR does not require advance knowledge of which variable drives the NMAR mechanism and is the recommended strategy whenever dependent nonresponse between the census and PES is a concern.

The three applications in Section~11 illustrate that the same mathematical structure accommodates substantively different comparison problems, and that the choice of paradigm is itself analytically important. PoE versus UR is naturally a Paradigm I problem, since the two methods cannot be run simultaneously at full scale. Imputation method comparison can use either paradigm depending on data availability. The comparison of field collection against administrative data sources is the natural home for Paradigm II: a national statistical office can use a census as the benchmark and evaluate administrative records against it retrospectively---building a rigorous, cost-effective evidence base for future census design decisions without the expense of a new large-scale validation survey.

In each application, the key analytical choices are: (i) which paradigm is feasible given available data; (ii) what serves as the reference $T_i$ and how reliably it can be established; and (iii) which subgroups are likely to drive differences between the methods. The framework does not resolve these choices, but it ensures that whichever choices are made, the resulting comparison is grounded in consistent, transparent, and well-defined statistical quantities.

\bibliographystyle{plainnat}

\end{document}